# Exploring Native Atomic Defects in NiTe$_2$

*Wen-Xiao Wang[1], Kaihui Li[1], Xiaoshan Dong[1], Hao Xie[1], Jinglan Qiu[2], Chunqiang Xu[5], Kai Liu[1], Juntao Song[1], Yi-Wen Wei[3,4,\*], Ke-Ke Bai[1,\*] Xiaofeng Xu[5,\*], Ying Liu[1]*

[1] College of Physics and Hebei Advanced Thin Films Laboratory, Hebei Normal University, Shijiazhuang 050024, China

[2] College of Physics and Hebei Key Laboratory of Photophysics Research and Application, Hebei Normal University, Shijiazhuang, Hebei 050024, China

[3] School of Applied Science, Beijing Information Science and Technology University, Beijing 100192, China

[4] Songshan Lake Materials Laboratory, Dongguan, Guangdong 523808, China

[5] Key Laboratory of Quantum Precision Measurement of Zhejiang Province, Department of Applied Physics, Zhejiang University of Technology, Hangzhou 310014, China

Nickel ditelluride (NiTe$_2$), a new discovered type-II Dirac semimetal whose Dirac node lies close to its Fermi level, is expected to exhibit exotic phenomena including Lifshitz transition and superconductivity. As we know, defects are inevitable for transition metal dichalcogenides and have significant impacts on the optical and electronic properties. However, the systematic study of defects in NiTe$_2$ is still lack. Here, by using high-resolution scanning tunneling microscopy combined with first-principles calculations, the point defects including the vacancy, intercalation and antisite defects in NiTe$_2$ are systematically investigated. We identified five main types of native defects and revealed that the growth condition could affect the type of native defects. By controlling the ratio of ingredient during synthesis, the types of point defects are expected to be manipulated, especially antisite defects. Additionally, we find native defects could slightly dope the topological surface states. Our results provide a facile way to manipulate defects for future optimizing the electronic properties of NiTe$_2$ and other related materials.

Transition metal dichalcogenides (TMDCs), combined with van der Waals force between layers, are particularly desirable not only for fundamental research but also for technological development[1-2]. Among TMDC material, there is a kind of type-II Dirac semimetal, for example, PtSe$_2$, PtTe$_2$, PdSe$_2$ and PdTe$_2$. With novel band topology and Lorentz-violating type-II Dirac fermions, they have many intriguing physical properties such as nonsaturating linear magnetoresistance, a nontrivial Berry phase for the light mass carriers, topological surface states with superconductivity and so on[3-9]. Compared with other similar material, the type-II Dirac nodes of NiTe$_2$ is closer to the Fermi energy due to the significant contribution of Ni d orbital states[10-12]. It's expected to provide an advantageous platform to study the topological properties of type-II Dirac semimetals[10, 12]. In recent reports, NiTe$_2$ is theoretically predicted to have superconductivity, Lifshitz transition and coexistence of type-I and type-II Dirac cones under pressure and strain[13-15]. NiTe$_2$ also has the huge potential of application in THz photonics and optoelectronics with high mobility and broadband fast response in the high frequency region [16]. Besides, the property of NiTe$_2$ is stable being exposed to air for years[17-18]. All these excellent properties make NiTe$_2$ a candidate material for future high performance device.

As we all know, TMDCs are enriched with a variety of intrinsic defects, such as boundary, vacancies, adatoms, and substitutional impurities[19-22]. The presence of these defects leads to a significant impact on the properties of TMDCs[23-28], such as introducing magnetic property in non-magnetic TMDCs, enhancing the electrochemical activity, tuning the electronic and optical properties and so on[29-35]. For example, recent reports have shown that Pt vacancies in PtSe$_2$ can introduce the surface magnetism, leading to high magnetoresistance value in the device [36]. In terms of the electrochemical activity, Li et al.[37]

revealed S-vacancies can activate and optimize the catalytic reactivity of $MoS_2$ in hydrogen evolution reaction by tuning the hydrogen adsorption free energy. As the result, they achieved the highest activity so far. For the electronic properties, Qiu *et al.*[28] reported that defects such as sulphur vacancies and grain boundaries in $MoS_2$ can significantly reduce the carrier-density. The defect engineering attracts so much extensive study that has become a branch to tune the physical and chemical properties for realization of novel features as well as high-performance devices. However, as one of TMDC, the structural defects in $NiTe_2$ have not been explored so far. Understanding the basic structure and the electronic properties of defects is critical for the future development of $NiTe_2$-based application in electronics, optoelectronics and catalysis.

Here, we focus on the structure identification and the electronic property of native point defects in $NiTe_2$ with scanning tunneling microscopy (STM) and spectroscopy (STS), which are powerful tools to study the local effect up to atomic level[38]. Supported by density functional theory (DFT) calculations, we identified five main types of native defects. Among all the native point defects, antisite defect is predominate, which is similar to the $MoS_2$ synthesized by PVD. Meantime, we find the growth condition could affect the type of defects. Through tuning the stoichiometry of ingredients during synthesis, the point defects are expected to be manipulated, especially antisite defects. What's more, the STS spectra show that high density of defects can slightly dope the surface states. Our results provide sufficient experimental data on the native point defects in $NiTe_2$ and provide the evidence for tuning the defects of $NiTe_2$ and other related TMD materials.

The single crystals of $NiTe_2$ were synthesized in Te solution. Accurately weighed amounts of high purity nickel powder and tellurium ingots were mixed thoroughly with a molar ratio of 1:8 in a glovebox and then sealed in an evacuated quartz tube. This quartz tube was then heated to 950 °C quickly in a sintering furnace and kept at this temperature for 48 h, before being slowly cooled down to 500 °C(3 °C/h), and finally being quenched into cold water. The excess amount of Te was centrifuged at 500 °C.

The samples were cleaved in situ for angle-resolved photoemission spectroscopy (ARPES) measurements. The background pressure is better than $5\times10^{-11}$ mbar. ARPES measurements were performed at 10 K using Scienta Omicron company DA30L with helium lamp (21.2 eV).

STM measurement was completed at 77K with the back pressure better than $3\times10^{-11}$ mbar with Createc STM system. The STM tip was made of tungsten wire electrochemically etched by NaOH solution and cleaned by thermal treatment. The Lateral dimensions observed in the STM images were calibrated using a Si (111)-(7×7) lattice and the STS spectra were calibrated using an Ag (111) surface. The STS spectra, i.e., the dI/dV-V curve, was carried out with a standard lock-in technique using a 661 Hz alternating current modulation with an amplitude of 10 mV.

The spin-polarized density functional theory (DFT) calculations were performed with the Vienna ab initio simulation package (VASP)[39-40], using the projector augmented wave (PAW) potentials[41-42]. The electron-electron exchange correlation was approximated by the generalized gradient functional of Perdew-Burke-Ernzerhof (PBE)[43]. We used a vacuum space of 1.5 nm along the z direction. A cutoff energy of 400 eV was used together with a Gamma centered 8×8×1 k-points grid[44] for a (4×4) supercell containing 64 carbon atoms of bilayer $NiTe_2$. The van der Waals interactions were described via the DFT-D2 method of Grimme[45]. Energy barriers were determined using the climbing-image nudged elastic band (CI-NEB) method[46-47]. STM images were simulated using the Tersoff-Hamann formalism with a (8×8) supercell (256 atoms) of bilayer $NiTe_2$.

Figures. 1(a)-1(b) shows the unit cell of 1T $NiTe_2$, which consists of repetitive slabs of three atomic sublayers (Te1−Ni−Te2). The orange balls indicate top Te sublayer and yellow balls indicate bottom Te sublayer. Figure. 1(c) shows a typical atomic-resolved STM image of $NiTe_2$ surface *in situ* cleaving along (001) surface, revealing hexagonally arranged structure. According to the atomic structure of 1T $NiTe_2$, each bright protrusion represents an individual Te atom at the topmost Te plane. The distance of the nearest bright

protrusions is $3.75 \pm 0.05$ Å, which agrees quite well with the lattice spacing of earlier report[12]. The ARPES results show a metal-type energy bands well reproduced by the calculated band structures indicated by red dashed lines. The white arrow shows two topologically protected surface states crossing the Fermi level which is spin polarized illustrated previously[10-11, 48]. More band structure information results are seen in SI. The ARPES results validate the high quality of sample including the clean and monocrystalline property.

The TMDCs always have various defects during the synthetic process. We also observed protrusions or depressions on the surface of NiTe$_2$ shown in Figures. 2 and S1 indicating the existence of native defects. The up two images without hollow dots in Figure. 2 show the same area at positive (empty states) and negative bias (filled states), respectively. Most of them present trigonal geometry due to the symmetry of the NiTe$_2$ crystal. What's more, the defects appear different topography depend on sample bias, especially positive and negative bias shown in Figures. 2 and S2. For convenience, we denote these defects as type A, B, C, D, E shown in Figure. 2. The statistical data including four samples of all the defects is presented in the column graph in Figure. 2(d). It's easily to recognize that C defect is the predominant point defect among all defects. The density of C defects reaches $6.4 \times 10^{12}/cm^2$. Such a defect concentration is surprisingly high as same as the MoS$_2$ whose defects largely modify its electronic structure[19]. So it is critical to study the defects here.

In order to understand the structure of defects, we scanned five types of distinct defects at atomic level shown in Figure. 3. From the representative atomic image in Figure. 3(a), we can identify one dark-hole site at the surface both in positive and negative bias. What's more, the dark site keeps the same topography at a series sample bias. This means the lack of one Te atom in the periodic row of Te atoms, labeled as $V_{Te1}$. The vacancy defect Te or Se at surface in other topological insulators and TMD materials also show the same topography, such as Bi$_2$Se$_3$, Bi$_2$Te$_3$, PtSe$_2$, PtTe$_2$ and so on[49-55]. For accurately identifying the defect A, DFT simulation is carried out shown in right panel of Figure. 3(a). The integrated local density of states (LDOS) images well reproduce the experimental results. Besides, there are also some vacancy defects with brighter edge, shown in Figures. 2(c) and S3(b)-3(c). These defects may be the $V_{Te1}$ combined with point defects at adjacent atom site. A few types of combined defects are shown in Figure. S4.

The surface atoms of both type B and C are continuous at various biases, shown in Figures. 3(b) and (c). So we can judge the defects locate under topmost Te sublayer. They both show a bright 1×1 triangle especially at negative bias. Due to the 3-fold symmetry of NiTe$_2$, one atomic defect will affect the three nearest sites. The smallest 1×1 triangle means a single point defect. Following, we will demonstrate that the type-B and C are identified as $V_{Ni}$ (Ni vacancy of the first layer) and $Te_{Ni}$ (substituted Te atom at Ni sublayer) defect, respectively.

The DFT simulations of vacancy and antisite defects locate at Ni sublayer which is under first Te atom layer shown in right panel of Figures. 3(b) and 3(d). Even though the simulations of $V_{Ni}$ and $Te_{Ni}$ both show three bright protrusions, $V_{Ni}$ shows small expansion of the in-plane lattice constant in the triangular protrusions due to the deformation after missing Ni-Te bond illustrated in Figure. S6. In experiment, type B shows triangular protrusions with distance 0.55±0.1nm, larger than 1×1 triangle which consistent with the simulations of $V_{Ni}$. So $V_{Ni}$ fit much better for STM image of type B, illustrated in Figure. 3(b). For type C, three atoms nearest to 1×1 protrusions are brighter at positive bias comparing with negative bias in experiment, well reproduced by the simulations of $Te_{Ni}$ defect. Besides, the brighter protrusions at positive than negative bias for $Te_{Ni}$ defect, indicate the electron-acceptor like behavior of $Te_{Ni}$.

Typically, the deeper the point defects locate, the weaker and wider the tunneling signal will be. Type D defect show protrusions with a larger 2×2 triangle, compared with $V_{Ni}$ and $Te_{Ni}$ defect, shown in Figure. 3(d). So D defect is supposed to locate deeper than $V_{Ni}$ and $Te_{Ni}$ defect. We performed DFT calculations for three possible atomic structures: Te vacancy of the bottom Te sublayer ($V_{Te2}$) in the first layer, intercalated Te atom ($Te_i$) or intercalated Ni atom ($Ni_i$) between the first and second layer. According to the simulations (see Figure. S6), $V_{Te2}$ also show a 2×2 triangle. However the 2×2 triangle of $V_{Te2}$ show the different atom direction with D defect (see Figure. S6). According to the simulations of $Te_i$ and $Ni_i$, the $Te_i$ simulation agrees well with the STM results of type-D shown in Figure. 3.

There is another type which is difficult to identify due to the very weak signal such as E defects. It shows a weak large 4×4 triangle at negative bias as illustrated in Figure. 3(d). Obviously, E defect lies deeper than all above defects. From the atomic direction of central triangle of E defect shown in Figure. 3(e), we can identify it locates at Ni site of the second layer $NiTe_2$. The simulate images of vacancy, antisite at Ni site of second layer $NiTe_2$ are shown in Figures. 3(d) and S7. The simulated $V_{Ni,2}$ defect shows slightly brighter 4×4 protrusions at negative bias, consistent well with the experimental results. Meanwhile, the central three atoms for the $V_{Ni,2}$, look slightly brighter both in STM and simulated images, further demonstrate the formation of $V_{Ni,2}$ defect.

Among all the defects, the antisite defect is predominant, which has close relation with the synthesis condition. The single crystals of $NiTe_2$ used here were synthesized with Ni powder and Te powder in a molar ratio of 1:8. In this Te rich condition, the excess Te is prone to replace Ni atoms to form antisite defects. Reflecting this fact, $Te_{Ni}$ defects have the lowest formation energy of 0.911eV, illustrated in Table 1. Such a feature is similar to the $MoS_2$ synthesized by PVD method[20]. What's more, the type of antisite in $NiTe_2$ could be affected by the ratio of ingredients during the synthesis, which is confirmed by formation energy calculations for different point defects, shown in Table 1. In Te-rich environment, $Te_{Ni}$ defects have the lowest formation energy among all point defects, while in Te-poor (Ni-rich) condition $Ni_{Te}$ defects have the smallest formation energy. Therefore, through controlling the ratio of chalcogens and transition metal element during the synthesis, the type of antisite defect ($Te_{Ni}$ and $Ni_{Te}$ antisite ) can be manipulated in $NiTe_2$.

For further understanding the effect of defects on electronic properties of $NiTe_2$, we carried out STS measurements on different regions and spots of the surface. Figure. 4 shows the STS spectra of pristine area and defects area. The STS of pristine area reveals two pronounced peaks, which are centered at approximately 10±5mV and −120±8mV labeled as α and β peak indicated by red line in Figure. 4(a). The intensity of α peak at Fermi level is much stronger than β peak indicated by STS spectra. Meanwhile, two similar peaks emerge at the angle-integrated ARPES spectrum in the black curve in Figure. 4(a). For clearly explaining the DOS peaks, we performed the simulated LDOS with 10 slab $NiTe_2$ model shown in Figure. 4(b), which also shows two peaks. The qualitative agreement between the experimental data and the calculation suggests that the two peaks are spin-polarized topological surface states, originated from van Hove singularities of the TSS band of $NiTe_2$. When carefully comparing the experimental result with the simulated one, there is still an obvious difference. The α TSS is at higher energy in experiment than simulation. Such a discrepancy may due to the abundant native defects in experiment. Above all, the STS experiment shows the energy gap between two spin polarized surface states is up to 130meV, which is larger than the previous experiment [11], further establishing $NiTe_2$ a promising for spintronic applications.

The STS spectra of defective area also display two surface states indicated in Figure. 4(c). Being robust against the disorders further suggests the topological nature of the surface states. Meanwhile, we performed STS measurements on the samples labeled as S and S' with different defect concentrations, shown in Figures. 4(d), (e) and S9. For S sample, α TSS is at 5±2meV above its Fermi level, while for higher defect density, α peak is at 17±3meV, which is obviously higher. Hence one can consider the point defects could dope the TSS of $NiTe_2$. And the higher the defect concentration is, the more α peak shifts. It is difficult to directly compare the TSS of defective area with calculated one due to the too large supercell when considering the reasonable concentration of defects. However, there are two obviously evidences, including the consistency of calculated DOS at native $NiTe_2$ with experimental one and the data at two concentrations in experiment, to show the defects could dope the TSS.

In summary, by using STM measurements and DFT calculations, we systematically studied five types of native defects in NiTe$_2$ at the atomic level, including single Te vacancy on top Te sublattice (V$_{Te}$), single Ni vacancy of first Ni sublayer (V$_{Ni}$), Te atoms substitute Ni atoms of first layer (Te$_{Ni}$), Te interstitial defects between the NiTe$_2$ layers (Te$_i$) and Ni vacancy of second NiTe$_2$ layer (V$_{Ni,2}$). Among all the defects, Te antisite is the dominant native defect. We find the type of defects has close relation with the growth condition. Through controlling the ratio of chalcogens and transition metal element during synthesis, the types of defect are expected to be tuned in crystal, especially antisite Te$_{Ni}$ and Ni$_{Te}$. Additionally, the point defects could slightly doped topological surface states. Our results provide sufficient experimental data on the native point defects in NiTe$_2$ and put forward a facial way to manipulate the defects, which would be crucial for future optimizing electronic properties of NiTe$_2$ and other related TMD materials.

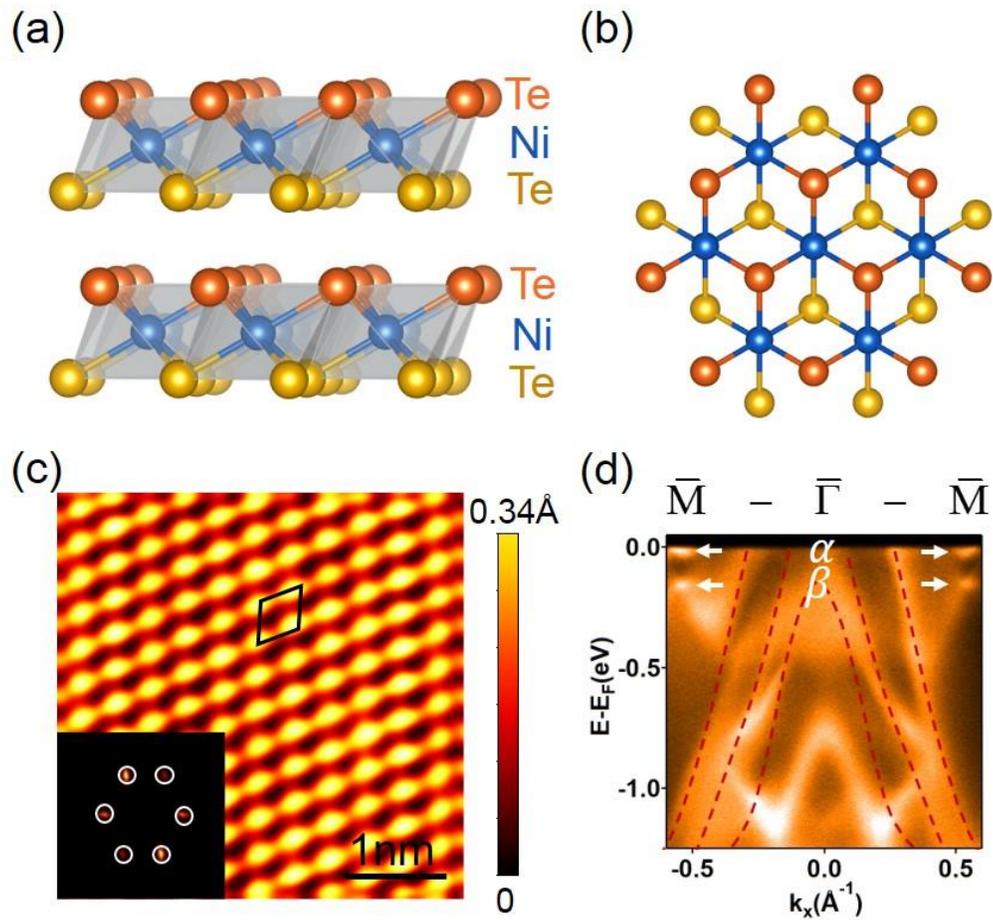

**Figure 1**. Characterization of structural and electronic structure of NiTe$_2$. (a)-(b) Side and Top view of the NiTe$_2$ crystal structure. Orange balls indicate upper Te atoms and yellow balls indicate bottom Te atoms. (c) High resolution atomic-resolved STM image (V=-0.7V, I=0.1nA). Inset is a fast Fourier transform diagram. The unit cell of NiTe$_2$ is marked by a black rhombus. (d) Experimental ARPES spectrum along the M- Γ-M direction in NiTe$_2$. The calculation band structure (red dotted line) is overlaid onto the image. The white arrows indicate the α and β energy band.

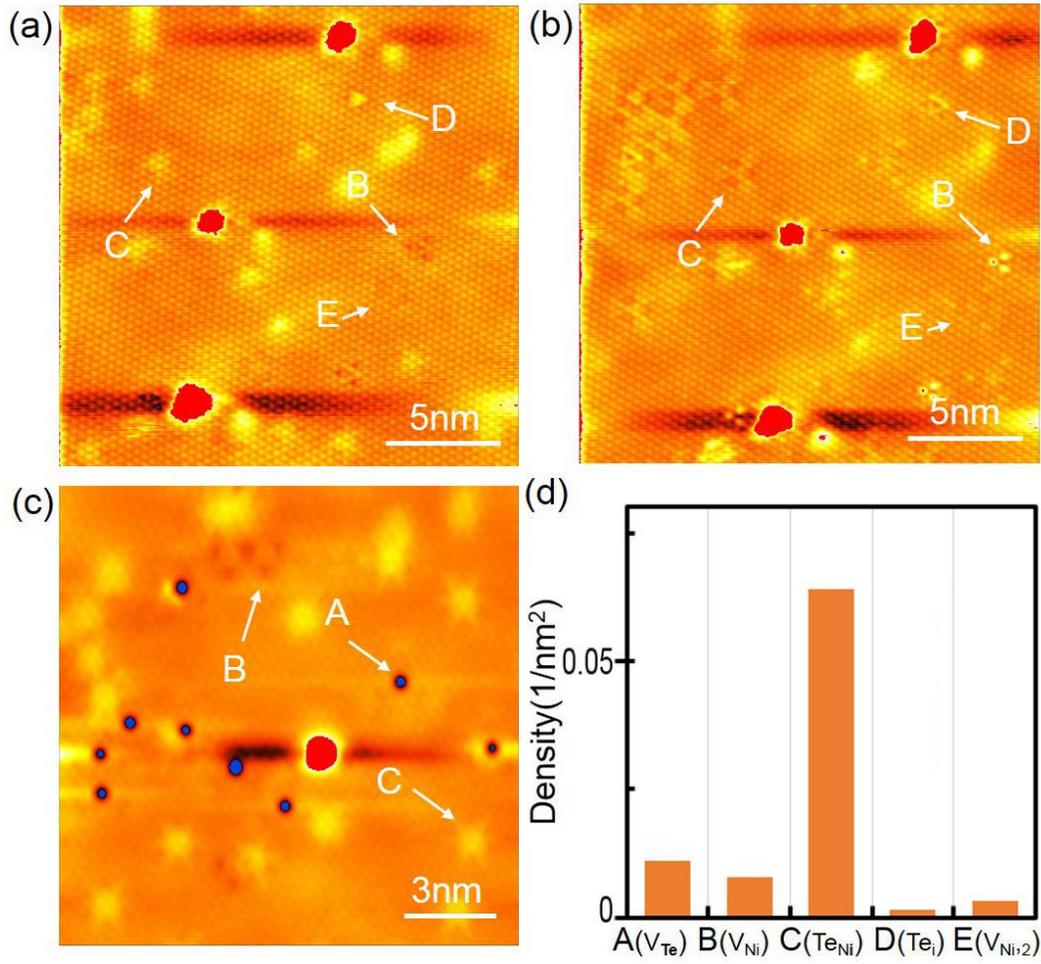

**Figure 2**. STM images of NiTe$_2$ surface at 77K. (a) Typical STM topography at Empty states (V=0.6V, =0.1nA). (b) Typical STM topography at Filled states (V=-0.6V, I=0.1nA). (c) Empty -states STM image of sample 2(V=1V, I =0.21nA). The white arrows mark the different types of defects. (d) Histograms of various point defects densities.

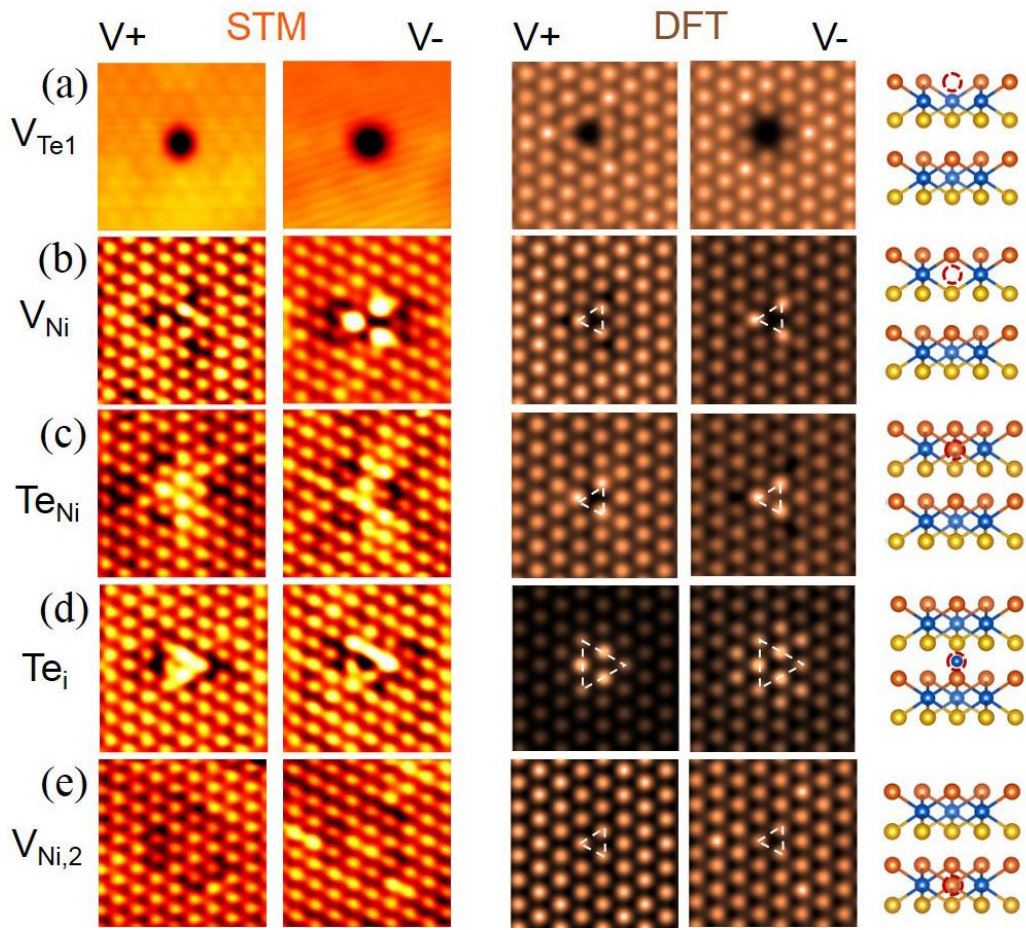

**Figure 3**. Typical atomically resolved STM topographies and DFT simulation of defects corresponding to A, B, C, D, E, respectively. Bias voltages and Setpoint: (a) V=±1V, I=0.21nA and (b)-(e) V =±0.6V, I=0.21nA. The bias voltage of all DFT images is ±0.6V. The size of images for STM and DFT simulation are 2.86nm×2.86nm and 2.70nm×2.70nm, respectively.

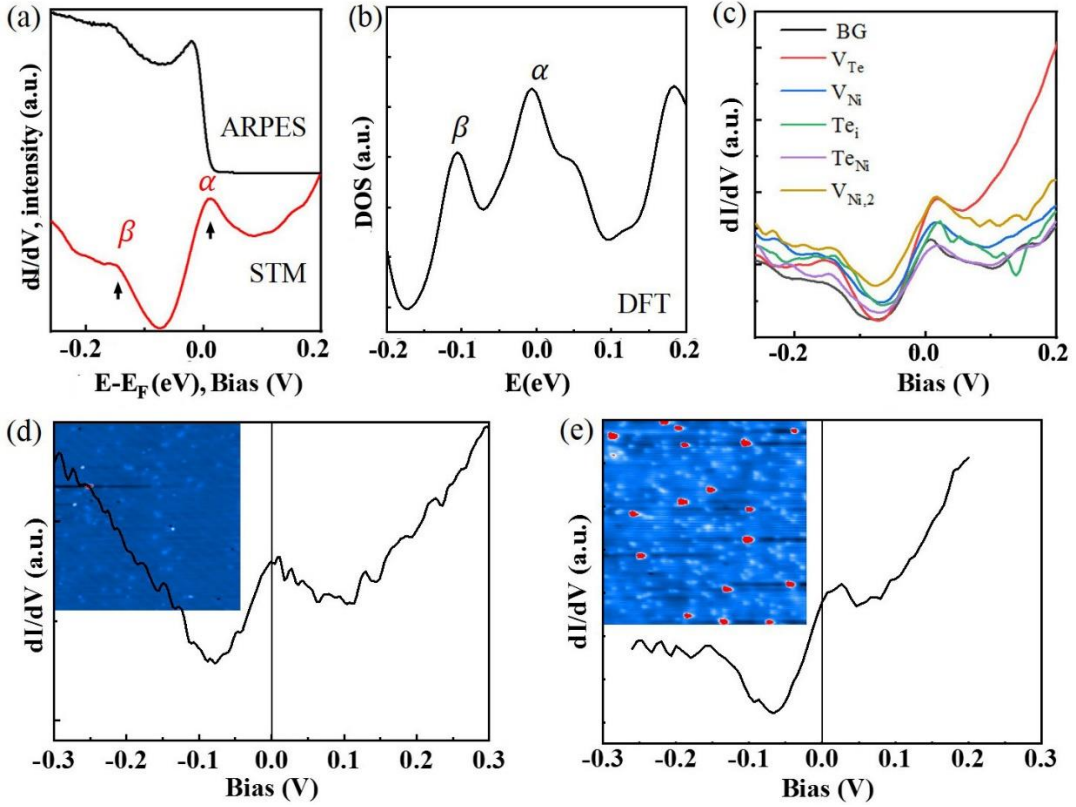

**Figure 4**. The topological surface states crossing Fermi level of NiTe$_2$. (a) ARPES spectra (top: Temperature=10 K, hν=21.2 eV, integrated along the Brillouin zone ΓM) and differential conductance spectra measured using STM (bottom: Temperature=77 K). (b) The DOS spectra of NiTe$_2$ (10 slab) based on DFT calculations. (c) Differential conductance spectra recorded on top of background, V$_{Te}$ (red line), V$_{Ni}$ (blue line)), Te$_i$, (green line), Te$_{Ni}$ (purple line) and V$_{Ni,2}$ (orange line), respectively. (d, e) The typical dI/dV-V curve obtained on the surface of samples with different density of defects. The STM images of the NiTe$_2$ surface shown in the inset, were measured with bias V$_{set}$ =1 V, setpoint I$_{set}$ =0.21nA in (d) and 0.64 V and 0.1 nA in (e).

Table 1. Formation energy $\Delta E(eV)$ of considered point defects.

| Defects | $V_{Te1}$ | $V_{Ni}$ | $Te_{Ni}$ | $Te_i$ | $Ni_{Te}$ | $V_{Ni,2}$ |
|---|---|---|---|---|---|---|
| **Te-rich** | 1.440 | 1.256 | 0.911 | 2.705 | 2.080 | 1.256 |
| **Ni-rich** | 1.115 | 1.904 | 1.883 | 3.030 | 1.107 | 1.904 |

**Supporting Information**

More results about the simulation of defects and more band structures in experiment (PDF)

Email: ywweiphy@foxmail.com
Email: kekebai@hebtu.edu.cn
Email: phyxxf@hotmail.com

The authors declare no competing financial interest.

W. Wang acknowledges the support from the National Natural Science Foundation of China (Grant Nos. 11904076, 21114044), Natural Science Foundation of Hebei (Grant No.A2019205313), the Key Program of Natural Science Foundation of Hebei Province (Grant No. A2021205024) and Science Foundation of Hebei Normal University (Grant No. L2019B10). Y. Wei was supported by Guangdong Basic and Applied Basic Research Foundation (Grant No. 2020A1515111196). X. Xu acknowledges the support from the National Natural Science Foundation of China (Grant No. 11974061).